\documentclass[a4paper,fleqn,usenatbib]{mnras}

\usepackage{mathptmx}

\usepackage[T1]{fontenc}
\usepackage{ae,aecompl}

\usepackage{amsmath}	
\usepackage{amssymb}	
\usepackage{graphicx,wrapfig,lipsum}
\usepackage{float}
\usepackage{multirow}
\usepackage{caption}
\usepackage{subcaption}
\usepackage{mwe,tikz}
\usepackage{array}
\usepackage{footnote}
\usepackage{threeparttable}
\usepackage{rotating} 
\usepackage{xcolor}

\title[Fast Photometric Variability in IC~348]{Fast Photometric Variability of Very Low Mass Stars in IC 348: Detection of Superflare in an M-dwarf}

\author[Ghosh et al.]{Samrat Ghosh$^1$\thanks{samrat687@gmail.com}
Soumen Mondal$^1$,
Somnath Dutta$^2$,
Ramkrishna Das$^1$,
Santosh Joshi$^3$,
\newauthor
Sneh Lata$^3$,
Dhrimadri Khata$^1$,
Alik Panja$^1$
\vspace{0.5cm}\\
\normalsize $^1$ S. N. Bose National Centre for Basic Sciences, Salt Lake, Kolkata 700106, India;\\
\normalsize $^2$ Institute of Astronomy and Astrophysics, Academia Sinica, Taipei 10617, Taiwan;\\ \normalsize  $^3$ Aryabhatta Research Institute of Observational Sciences (ARIES), Manora Peak, Nainital-263002, India }

\date{Accepted XXX. Received YYY; in original form ZZZ}

\pubyear{201-}

\begin{document}
\label{firstpage}
\pagerange{\pageref{firstpage}--\pageref{lastpage}}
\maketitle

\begin{abstract}

We present here optical {\it $I$}-band photometric variability study down to $\simeq$ 19 mag of a young ($\sim$2-3 Myr) star-forming region IC~348 in the Perseus molecular cloud. We aim to explore the fast rotation (in the time-scales of hours) in Very Low Mass stars (VLMs) including Brown Dwarfs (BDs). From a sample of 177 light-curves using our new {\it $I$}-band observations, we detect new photometric variability in 22 young M-dwarfs including 6 BDs, which are bonafide members in IC~348 and well-characterized in the spectral type of M-dwarfs. Out of 22 variables, 11 M dwarfs including one BD show hour-scale periodic variability in the period range 3.5 - 11 hours and rest are aperiodic in nature. Interestingly, an optical flare is detected in a young M2.75 dwarf in one night data on 20 December 2016. From the flare light curve, we estimate the emitted flared energy of 1.48 $\times$ 10$^{35}$ ergs. The observed flared energy with an uncertainty of tens of per cent is close to the super-flare range ($\sim$  10$^{34}$ ergs), which is rarely observed in active M dwarfs. 
\end{abstract}




\section{Introduction}

 The very low-mass stars (VLMSs) refer to the stellar and substellar objects with mass below 0.6 $M_\odot$ to planetary limit (0.013 $M_\odot$) (e.g. Allard et al. \citeauthor{Allard1997}), which include from spectral type of mid-K , M, L, T to the coolest known dwarf Y (Lodders \& Fegley \cite{Lodders2006}; Burgasser et al. \cite{Burgasser2006}; Cushing et al.\cite{Cushing2011}; Kirkpatrick et al. \cite{Kirkpatrick2012};  Allard et al.  \cite{Allard2012}). They extend from  the edge of the hydrogen-burning main sequence (MS; i.e., mid-K and early-M) to the deuterium-burning Brown Dwarfs (BDs; 80 - 13 $M_J$) with transition at M6 spectral type at the young age of IC~348, which is less than 5 Myrs (Chabrier \& Baraffe \citeauthor{Chabrier2000b}, Luhman et al. \cite{Luhman2003}; \cite{Luhman2005};  \cite{Luhman2016}, Spiegel et al. \cite{Spiegel2011}, Zhang et al. \cite{Zhang2017}). The VLMs are the most common objects in our Galaxy, and they span over a wide range from young, metal-rich M dwarfs in open clusters (Simons \& Becklin \cite{Simons1992}, Zapatero Osorio et al. \cite{Zapatero1996} and references therein; Zhang et al. \cite{Zhang2018}, Caballero et al. \cite{Caballero2019}), and the galactic disk (Gliese \& Jahreiss \cite{Gliese1991}) to the oldest metal-poor sub-dwarfs of the galactic field (Green \& Margon \cite{Green1994}, Kirkpatrick et al. \cite{Kirkpatrick2006}, Cruz et al. \cite{Cruz2009}, Dupuy and Liu \cite{Dupuy2012}; \cite{Dupuy2017}, Zhang et al. \cite{Zhang2019}, Carmichael et al. \cite{Carmichael2020}) and globular clusters (Richer et al. \cite{Richer1995}, Renzini et al. \cite{Renzini1996}, Dieball et al. \cite{Dieball2016}). Their studies on diverse environment, therefore, enhance our knowledge on the reveal substantial information about the dynamical and chemical evolution of the galaxy (Schmidt et al. \cite{Schmidt2016}).

Several studies were performed to explore VLMs characteristics in star-forming regions (SFRs), such as Taurus (Luhman et al. \cite{Luhman2009} , Esplin et al. \cite{Esplin2017}), Perseus (IC~348 : Luhman et al. \cite{Luhman2016}; NGC1333 : Scholz et al. \cite{Scholz2012}), Chamaeleon (Esplin et al. \cite{Esplin2017}), Orion (ONC: Caballero et al. \cite{Caballero2007}; $\sigma$ Ori.: Zapatero Osorio et al. \cite{Zapatero2000}), Ophiuchus (Oliveira et al. \cite{Oliveira2012}), Lupus (Mu{\v{z}}ic et al. \cite{Muzic2014}), Upper Scorpio (Lodieu et al. \cite{Lodieu2018}) etc. These young dwarfs rotate with relatively faster time-scales than the objects in the galactic field (Cody \cite{Cody2010}, Scholz et al. \cite{Scholz2011}), and the short rotation periods mean that one or more rotations can be observed during a single night observation. Ground-based 1- 2m class telescope facilities are good enough to explore the photometric variability of this low-mass faint regime in near-by SFRs of our Galaxy.

 Stellar flares are often observed on M dwarfs (West et al. \cite{West2008}; Hilton et al. \cite{Hilton2010}; Pineda et al. \cite{Pineda2013}), with an a wide range of flare energies, from $E \sim 10^{26}$ ergs up to  $10^{35} - 10^{36}$ ergs (Kowalski et al. \cite{Kowalski2010}; Davenport \cite{Davenport2016}; Schmidt et al. \cite{Schmidt2019}). Flares are basically multi-wavelength outbursts originated by the reconnection events of magnetic field lines at the stellar surfaces, which appears in sharp flux increase followed by a more steady exponential decrease within a few minutes up to a few hours (France et al. \cite{France2013}; Davenport et al. \cite{Davenport2014}; Jones \& West \cite{Jones2016}; Mart{\'i}nez et al. \cite{Martinez2019}). M dwarfs rotation can induce a dynamo causing magnetic activity in the photosphere (Goulding et al. \cite{Goulding2012}). Moreover, optical wavelengths are proven to be an essential resource for flare observations in the Kepler mission (Borucki et al. \cite{Borucki2010}) and its extension into K2 (Howell et al.\cite{ Howell2014}).

Photometric variability studies of VLMs is an important tool to probe the physical nature of their atmospheres. Most BDs (70 \% L dwarfs; Rockenfeller et al. \cite{Rockenfeller2006}) are found to be variable in broadband photometry. Numerous studies of such variability studies are attempted earlier in the optical and infrared wavelengths (e.g.,  Tinney \& Tolley \cite{Tinney1999}, Bailer-Jones \& Mundt \cite{Bailer1999}, Gelino et al. \cite{Gelino2002},  Koen et al. \cite{Koen2005}, Morales-Calderon et al. \cite{Morales2006}, Clarke et al. \cite{Clarke2008},  Radigan et al. \cite{Radigan2014b}, Metchev et al. \cite{Metchev2015}, Apai et al. \cite{Apai2017}). It was suggested that weather-like patterns in VLMs and BDs might result in rotation-induced variability. Photometric variability in dwarf is due to the presence of surface features like magnetic spots (due to strong magnetic fields) or dust clouds or binary companion, which cause optical modulation as it rotates (Lew et al. \cite{Lew2016}, Kostov \& Apai \cite{Kostov2013}). Typical measured {\it{vsini}} values of L-dwarfs are in the range 10 to 60 km/s, which corresponds to rotation periods of 2-12 hours (Mohanty \& Basri \cite{Mohanty2003}). BDs, being rapid rotators, having a period of few hours to days (Herbst \cite{Herbst2000}, Crossfield \cite{Crossfield2014}), the variability in those dwarfs could be measured within a few nights of photometric monitoring using small to moderate-sized telescopes. Rotational modulation in the light curves of such dwarfs provides the period of rotation of the object. Also, rotating such dwarfs may transfer momentum to circumstellar disks via interacting with the magnetic field (disk locking;  Herbst \cite{Herbst2000}). Palla \& Baraffe (\cite{Palla_Baraffe2005}) proposes another hypothesis that deuterium burning VLMs (greater than M4) and BDs with a mass range from 0.02 - 0.1$M_\odot$ would exhibit radial pulsation due to the conversion of their nuclear energy to kinetic energy resulting in oscillation with a period ranging from 1 to 4 hours.

\begin{figure*}
\begin{center}
\includegraphics[trim={0.75cm 5.7cm 0 0.55cm}, height=4.in]{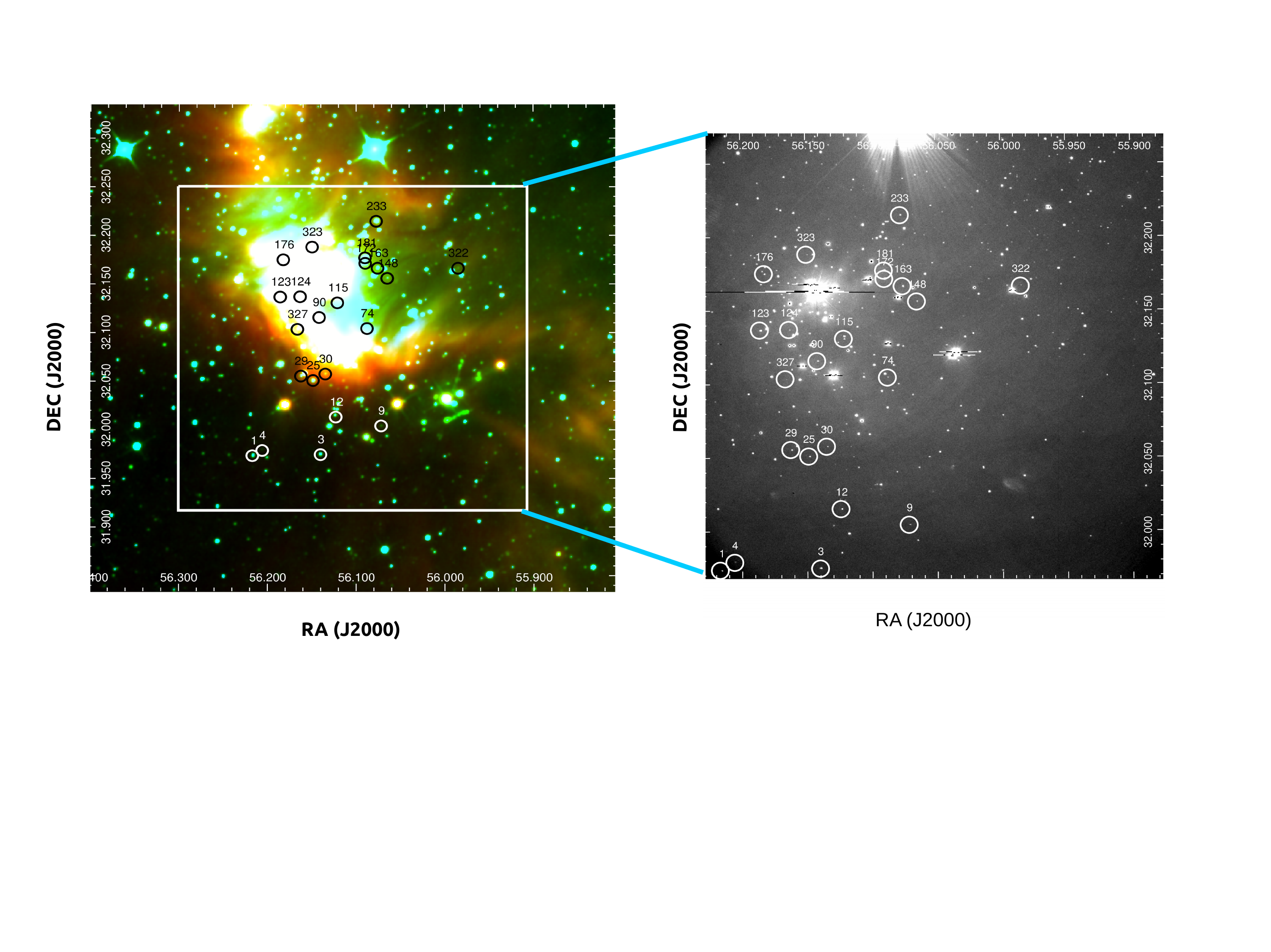}
\caption{{Left: Colour composite image of IC~348 generated with 2MASS K$_s$ (blue), WISE 3.4 $\mu$m (green), WISE 4.6 $\mu$m (red). Combined monitoring observations were performed in the marked square-region. Right: An image of {\it $I$}-band observations taken on 18 December 2016 using 1.3-m DFOT. The variable candidates are marked by open circles in both figures.}}
\label{ic348}
\end{center}
\end{figure*}

Our target list was a well-characterized sample of VLMs in late M spectral type (see Figure \ref{ic348}, for spatial distribution), which are spectroscopically confirmed bonafide members in IC 348 known from the literature (Scholz et al.\cite{Scholz1999}, Luhman et al.\cite{Luhman2003}, Muench et al.\cite{Muench2007}). IC 348 is a young (1 - 3 Myr; Herbig \cite{Herbig1998}, Muench et al.\cite{Muench2003}) and near-by (310 pc; Luhman et al.\cite{Luhman2003}) star-forming region in Perseus molecular cloud (Luhman et al. \cite{Luhman2003}, D'Antona \& Mazzitelli \cite{DM94}). The IC~348 region is well studied with spectroscopic and photometric measurements in IR, Optical and X-Rays (Lada \& Lada \cite{Lada95}; Luhman et al.\cite{Luhman1998};  Preibisch \& Zinnecker \cite{Preibisch2001}; Carpenter et al.\cite{Carpenter2002};  Muench \cite{Muench2003}; Lada et al. \cite{Lada2006}; Muench et al. \cite{Muench2007}; Luhman et al.\cite{Luhman2003}; Cohen et al.\cite{Cohen2004}; Luhman et al.\cite{Luhman2005}, Dahm \cite{Dahm2008}; Alexander et al.\cite{Alexander2012}; Esplin et al.\cite{Esplin2017}). Because of IC 348 cluster's intermediate star density  ($ \rho \sim 100-500~M_{\odot} pc^{-3}$; Parker \& Oliveira (\cite{Parker2017})), it has enough stars ($\sim$500; Luhman et al. \cite{Luhman2016}) to detect large populations of low mass objects ($N(>M6.5)/N(<M6.5)= \text{0.188}^{+0.025}_{-0.02}$ ; Luhman et al. \cite{Luhman2016}) in 10-20 arcmin Field of View and light from over-populating bright stars doesn't pollute the faint low-mass sources (Nordhagen et al. \cite{Nordhagen2006}). D'Antona \& Mazzitelli (\cite{DM94}) and Baraffe et al. (\cite{Baraffe2003}) suggested from their models that the hydrogen-burning mass limit is M6 with consistent temperature range at ages $\le$ 10 Myr. Previous surveys found a large population of candidate BDs in IC 348 (Luhman et al. \cite{Luhman2003}, Luhman et al. \cite{Luhman2005}).It motivates us to use IC~348  for variability studies of M dwarfs including BDs.

The paper is organized as follows: section \ref{sec_obs}  describes our observation log and section \ref{sec_data_red} describes the data reduction process in brief. In section \ref{sec_res}, we have the results of our work and discussed the results. We have summarised our work in section \ref{summary}.

\section{Observations}
\label{sec_obs}
Photometric data were obtained using the 1.3-m Devasthal Fast Optical Telescope (hereafter, 1.3-m DFOT) located at Devasthal, Nainital, India (Sagar et al. \cite{Sagar2011}). The ANDOR 2K$\times$2K CCD instrument was used for the optical observations. This instrument has a pixel size, of 13.5 $\mu$m. We use a Johnson-Cousin {\it I} filter with an unvignetted field of view (FoV) of 18 $\times$ 18 $arcmin^2$. The camera was operated at 1 MHz readout mode with an RMS (root mean squared) noise of 6.5 $e^-$ and a gain of 2.0 $e^-$/ADU. Images were taken in I-band with exposure of 300 and 400 seconds in the kinetic-mode (multiple frames) as well as in the single image mode. 
Data were also obtained from the 2-m Himalayan Chandra Telescope (hereafter, 2-m HCT), located at Hanle, Ladakh, India. The backend instrument Himalayan Faint Object Spectrograph and Camera (HFOSC), the $2K \times 2K$ part of the detector in $2K\times4K$ CCD having a pixel size of 15 $\mu$m and a pixel scale of 0.296''  is used for imaging observations (Prabhu \cite{Prabhu2014}). The FoV on $2K \times 2K$ part of CCD in the imaging mode is $10 \times 10 ~arcmin^2$, which is used for {\it I} band imaging observations. The log of observations is mentioned in Table I. \\ \\

\renewcommand{\tabcolsep}{5pt} 
\begin{table*}
\centering
\caption{\bf{Observational Log for IC~348}}
\label{log-table}
 \hspace*{-1.cm}
\begin{tabular}{cccccccccc}
\hline
\vspace*{.20cm}
Date & Object & Telescope & Instrument &  \begin{tabular}[c]{@{}c@{}}FoV\\ ($arcmin^2$)\end{tabular} & Filter & Exposure (sec) $\times$ N & \begin{tabular}[c]{@{}c@{}}Run Length\\ (hours)\end{tabular} & \begin{tabular}[c]{@{}c@{}}Average See-\\ ing (arcsec)\end{tabular} & \begin{tabular}[c]{@{}c@{}}Night\\ Condition\end{tabular} \\
\hline\\
18.12.2016 & IC~348	&	1.3-m DFOT & ANDOR~2K$\times$2K~CCD &\hfil 18$\times$18 & {\it $I$} & 300$\times$1,400$\times$28 &	 \hfil 3.77 & 2.1 & dark \\
19.12.2016 & IC~348	&	1.3-m DFOT & \hfil''	& \hfil '' & {\it $I$} &  400$\times$19  		& \hfil 1.85 & 2.4 &dark  \\
20.12.2016 & IC~348	&	1.3-m DFOT &\hfil''	& \hfil ''  & {\it $I$} & 400$\times$40	&   \hfil 6.13 & 2.6 & dark \\
27.10.2017 & IC~348  &	1.3-m DFOT & \hfil''	& \hfil ''  & {\it $I$} &  250$\times$3, 360$\times$1, 300$\times$35 &  \hfil 6.99 & 2.2 & dark \\
28.10.2017 & IC~348	&	1.3-m DFOT & \hfil''	& \hfil ''  & {\it $I$} & 300$\times$29, 360$\times$19 & \hfil  4.64 & 2.1 & dark \\
10.11.2017 & IC~348	&	1.3-m DFOT & \hfil''	& \hfil '' 	& {\it $I$} & 300$\times$45,360$\times$5 &  \hfil 6.16 & 2.3 & dark \\
14.11.2017 & IC~348	&	2-m HCT 	  & HFOSC~2K$\times$2K~CCD  	& \hfil 10$\times$10 & {\it $I$}&  300$\times$56 & \hfil  7.98 & 1.9 & dark \\
15.11.2017 & IC~348	&	2-m HCT 	  & \hfil''	& \hfil '' & {\it $I$} &  300$\times$18,360$\times$43	& \hfil 9.08 & 2.2 & dark \\\\ \hline

\end{tabular}

\end{table*}

\section{Data Reduction}
\label{sec_data_red}
\subsection{Data Analysis}

The raw images were reduced using standard packages in the {\sc iraf} software\footnote{Image Reduction and Analysis Facility ({\sc iraf}) is distributed by National Optical Astronomy Observatories (NOAO), USA (http://iraf.noao.edu/)} following {\it bias subtraction}, {\it flat field} corrections and {\it cosmic ray removal}. We first prepare a median stacked flat image and a median stacked bias image. Then the {\it{ccdprocess}} task is used to process all individual images and get flat and dark corrected images.  We have used the {\sc iraf}'s {\it daofind} (Stetson \cite{Stetson1992}) task to find sources in the frames. Manual removing and marking of sources using the {\it tvmark} task have also been done where the frames contain saturated sources (which are not required for our study) or undetected faint sources by the tasks. Very faint sources are missed by this task. Because sometimes if we lower our detection limits, then the task automatically detected some random background region nearer to brighter sources. To avoid detecting these regions we adjusted to an optimal lower threshold. And manually marked those very faint sources. The critical parameters like object detection threshold above local background, {\it threshold} are set between 3-4, the full-width at half-maximum of the point spread function, {\it fwhmpsf} is set between 5 to 8 (arcsec per pixel) as different telescopes have different values.

\subsection{Astrometry}

WCS (world coordinate system) coordinates of the detected stars are obtained using the 2MASS point source catalogue (Cutri et al. \cite{Cutri2003}) as references. A list with 25 sources is chosen from our frame, and their coordinate (RA \& DEC) in the 2MASS catalogue and pixel coordinates are matched. Then the {\it ccmap}\footnote{\href{https://iraf.net/irafhelp.php?val=images.imcoords\&help=Help+Page\&pkg=1}{package: images.imcoords}} task is implemented to find the plate solution of the image via the celestial coordinate and pixel coordinate. Using this solution, WCS coordinate is generated using {\it ccsetwcs{$^2$}} task. We obtain the astrometry accuracy of $~$0.3$\arcsec$.

\subsection{Aperture Photometry}

 We compiled a list of all M dwarfs of IC~348 available in the three catalogues mentioned above (Scholz et al. \cite{Scholz1999}, Luhman et al. \cite{Luhman2003}, Muench et al. \cite{Muench2007}), and tallied it against the detected sources in the frames. We found 177 M dwarf stars including brown dwarfs are in the catalogues. We used these as our object sample and proceed further. All 177 detected M-type sources, including brown dwarfs in IC~348 are chosen from the observed field using the available catalogue of IC~348 in the literature as mentioned before (Scholz et al.\cite{Scholz1999}, Luhman et al.\cite{Luhman2003}, Muench et al.\cite{Muench2007}). Aperture photometry using {\sc iraf}'s {\it phot} task is performed (1st run of phot) on the selected target sources as well as other unsaturated sources present in the frame by selecting radii from 1 to 25 pixels. We use this large range of radii for choosing the aperture with less error as an appropriate aperture is selected from the Growth Curve (instrumental magnitude vs Aperture plot) by visual inspection. The instrumental magnitudes of the sources don't change much with increasing the aperture after this "appropriate" aperture. We use this aperture in the 2nd run of the task {\it phot} to get the final magnitudes, which we use for differential photometry (sec \ref{diff_phot}). We choose sky annulus inner and outer radius outside of this "appropriate" aperture for the background subtraction. The standard stars have been used to calculate the zero-point for each night to get the calibrated magnitude. There are few brighter stars in the field which we did not consider because their counts were saturated or high enough with the long exposure (needed for having good signal-to-noise ratio for the faint sources) to be in the non-linear region of the CCD detector. 

\subsection{Differential Photometry}
\label{diff_phot}

Differential photometry is performed on all detected sources to get better light curves by removing the effects of atmospheric transparency and instrumental signatures. A time-series data on each source is obtained from the estimated magnitudes from all frames in our observing runs. We then apply the differential photometry on the reduced time-series data. The non-variable sources are chosen in such a way that their brightness/magnitudes are similar to the targeted object. After visual inspection of raw-light curves, we choose 20 likely non-variable stars and then create an average reference non-variable light curve. So, the advantage of using this average light curve is that it minimizes the local background effect, cosmic ray hits on individual data points. Then the only variation present in this time series data is the intrinsic variation which depends on the non-linear parameters like atmospheric conditions, airmass, instrumental parameters and which is embedded in every source including our targets and is unique in each frame. For each night of observations, we get a differential magnitude (source - average) on each point of the differential light curve,  which takes care of the data jump, extrinsic variability, etc. in different observing conditions. The differential light curves present only the intrinsic variation of the targets. Such a technique provides an effective way to detect and classify the variability(Mondal et al. \cite{Mondal2010}, Dutta et al. \cite{Dutta2018}, Dutta et al. \cite{Dutta2019}). The light curves of all detected sources, including the VLMs and BDs of interest, have been generated. We then combined all data points for each source covering a range of a year from 2016 to 2017 (see Table \ref{log-table}). We used the root-mean-squared (RMS) value of the light curves to select the objects with large RMS values implying significant peak-to-peak variation. We also visually checked all the individual-night light curves for periodic signals and any misinterpretation from the RMS method. This is a much-needed step as many low-amplitude variables might be missed in the RMS plot due to the high noise level. In this process of visual checking, a few variable sources from the RMS plot are excluded as their variability is false due to their terminal position in the CCD or bad or hot pixel. Similarly, a few sources with small peak-to-peak variation are also included as they are low-amplitude variables (and shows periodic light curve), which was masked by large error due to their faintness. 

\subsection{Periodogram Ananlysis}	
\label{periodogram}

Lomb-Scargle periodogram (LS periodogram; Lomb \cite{Lomb1976}; Scargle \cite{Scargle1982}) is computed using NASA Exoplanet Archive Periodogram Service\footnote{https://exoplanetarchive.ipac.caltech.edu/cgi-bin/Periodogram/nph-simpleupload} to find the significant periodic signals in the light curves and construct the phase light curves. It is a widely used algorithm in observational astronomy to find the periodic signals in an unevenly spaced time-series data. The LS periodogram uses a Fourier-like power spectrum estimator for the data to determine the period of oscillation. Results from the LS periodogram which were exactly or nearly equal to any of the lengths of the relevant observing runs (Table \ref{log-table} final column) are excluded as they are likely to be due to an alias caused by the gap in the observations. The CLEAN algorithm (Roberts et al. \cite{Roberts1987}) was not used in our data because the data is unevenly spaced whereas the CLEAN algorithm is based on classical FFT analysis which would only work if the data were evenly spaced. The Lomb-Scargle periodogram imitates the classical periodogram in the limit of evenly spaced data (VanderPlas \citeauthor{VanderPlas2018}). We checked each light-curve with Plavchan (Plavchan et al. \citeauthor{Plavchan2008}) and Box-fitting least square (Kovacs et al. \citeauthor{Kovacs2002}) periodogram algorithm to confirm the periodicity of a source. 
The Plavchan method is similar to the phase dispersion minimization (Stellingwerf \citeauthor{Stellingwerf1978}) algorithm where a periodic basis curve is computed from the data. This bin-less method uses box-car smoothed phased time-series for comparison with the phase curve folded with a trial period and find the best-matched curve.
The box-fitting least square (BLS) method fits the input data to periodic box functions instead of using sinusoids like LS method. The BLS periodogram is optimized for finding transit-shaped periodic signals in time-series data like transiting exoplanets or eclipsing binaries.
We also checked the phase curves for periodic signals using other available packages for cross-matching with the {\sc period04} software (Lenz \& Breger \cite{Lenz2005}) and {\sc starlink} software (Currie \cite{Currie2014}). The periods match well within an error of a few percents except a few faint and scattered light-curves.
All measured periods from our light curves converge from the different algorithms.

 The software/service used for this period calculation does not provide with an error except {\sc starlink} software that too very small. We have calculated the systemic error, $\delta f = 3 \sigma_N / 2TA \sqrt{N_0}$ (Horne \& Baliunas \cite{Horne1986}), where $\sigma_N^2$ is the variance of the noise in the data after the periodic signal is subtracted, T is the total span of data ($\sim$330 days), A is the signal amplitude and $N_0$ is the number of independent light-curve data points. Using this formula, we obtain the formal error of the order of $\sim$ 0.001 - 0.0001 h.

\section{Results and Discussion}
\label{sec_res}
\begin{figure*}
  \centering
    \hspace*{-.2in}
\includegraphics[ height=7.5in]{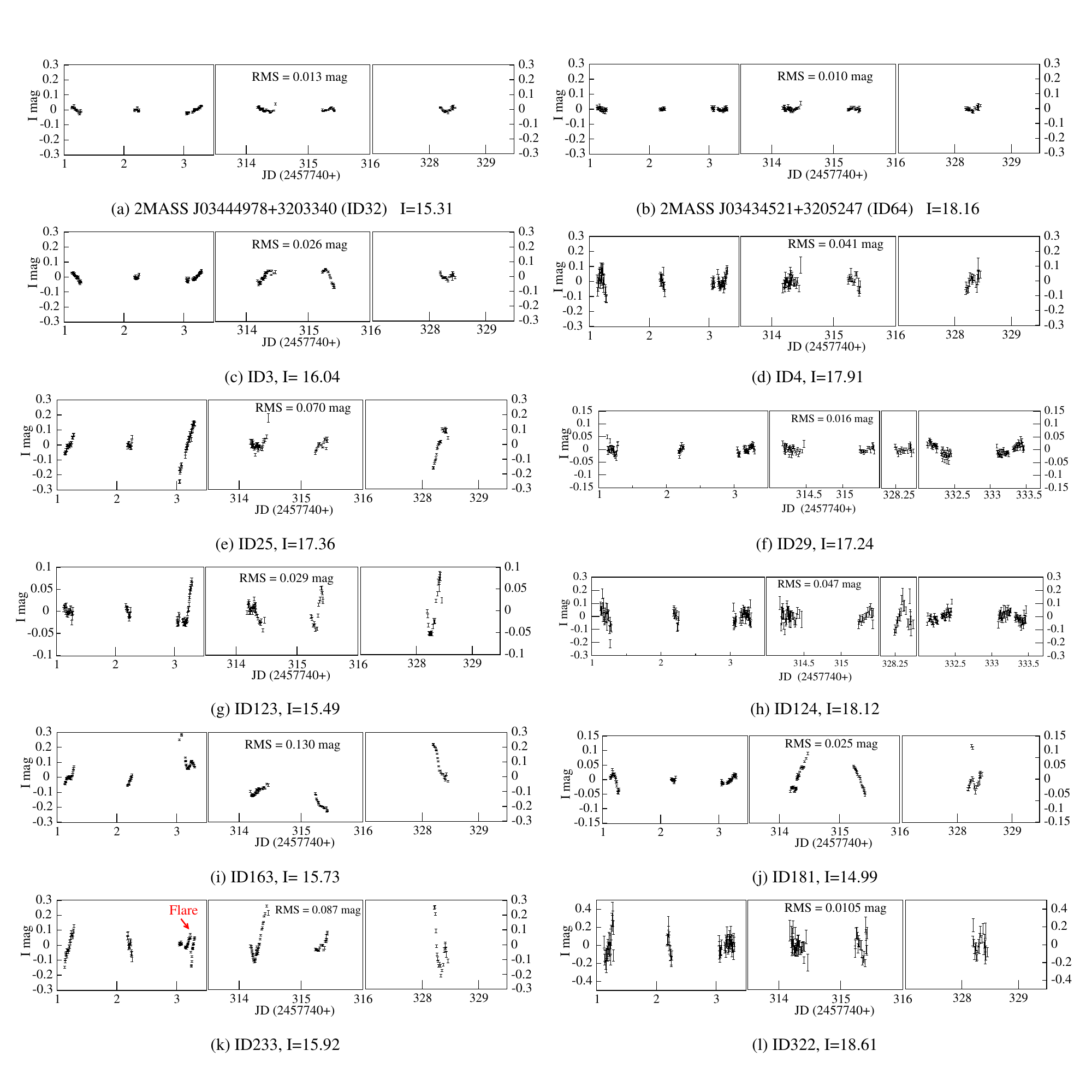}
\caption{{For example, a few observed light curves of variable stars including two non-variables in the 1st panel are shown.
 The x-axis is broken due to data gap in between observations. The y-axis is zero averaged {\it $\Delta$I} magnitude of the sources. The x-axis length is different in a few light curves and proportionate to the observing run-length (hours). The source ID and {\it $I$}-band magnitude is mentioned underneath each panel.}}
\label{lc0}

\end{figure*}

A time-series {\it $I$}-band photometric analysis was performed on the cluster members of IC~348. Using differential photometry and RMS plots (discussed in subsection \ref{diff_phot}) the intrinsic variation in the objects' light curves is unveiled. The light curves of a few variable objects are shown in Figure \ref{lc0}, which show significant variability. The light curves of two non-variables of similar brightness are also shown in Figure \ref{lc0} to judge the quality of the time-series data. A source is considered variable if their magnitude of variation (standard deviation in the light curve data - $\sigma$) is at least 3 times higher (3$\sigma$) than that of its reference sources of similar brightness. Few sources are loosely called a candidate variable if they show consistently $2\sigma$ or higher magnitude variation over all the observing nights. The details of the sources are provided in Table \ref{table2}. The summary of the results is shown in the same table.

\begin{figure*}
  \centering
  \hspace*{-.7in}
\includegraphics[height=7.in]{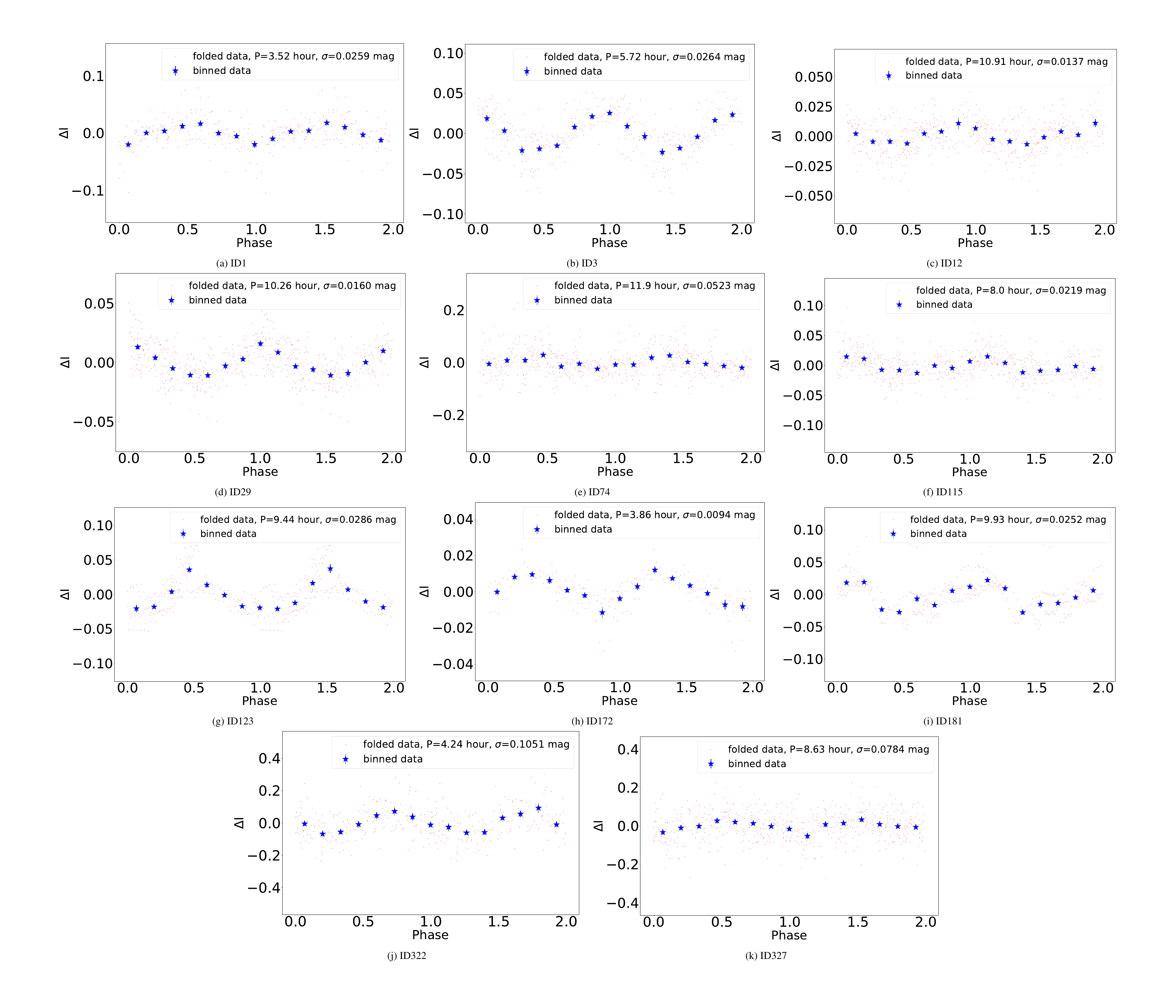}
\caption{Phase light curves of the periodic sources in red dots are shown here. The blue stars represent the 15-points binned data.}
\label{phase}

\end{figure*}

If any source gives a periodic signal in at least two of the three packages (see subsection \ref{periodogram}) within an error, then we call it a significant period. We choose only those significant periods which produced visually periodic phase curves and discarded other periods obtained from the periodogram computations. We folded the whole range of data with that period to compute the respective phase curves. We binned the light curves (15 points, corresponding to approximately 1.5 hrs) of each phase curve for better visualization of the periodicity in the data (see Figure \ref{phase}). We have included the sources which show periodicities at least in more than one night individually.

The estimated hour-scale periods of the identified variables computed from LS periodogram including other details are mentioned in Table \ref{table2}. From the 177 M dwarf's light curves with our new {\it $I$} band observations, we detect new photometric variability in 22 M dwarfs including 6 BDs. Out of these 22
variables, 11 of them including a BD show an hour-scale periodic variability in the period range 3.5 - 11 hours and rest are aperiodic. The estimated periods and amplitudes of variability are listed in Table \ref{table2}. The peak-to-peak variation is estimated from zero averaged light curves of the objects, and the value of RMS is given for aperiodic variables.

Figure \ref{phase} shows the phase light curves from time-series data folded with the period of the object as mentioned in Table \ref{table2}.

It is evident from the phase curves that the object ID~1 (an M3.5 dwarf) shows an apparent variability with a period of 3.52 hours. ID~3 (an M3.75 dwarf) is also known periodic variable with a period of 3.9 days (Flaherty et al. \cite{Flaherty2013}), but we detect a short-period of 5.72 hours from our data. From our observed large data cadence, we could not detect 3.9 days. Our study is focused on the short period of the M dwarfs based on the ground-based observations. ID~12 (an M6 dwarf) and ID~29 (an M5 dwarf) show small-amplitude ($\sim$ 15 mmag) with a periodicity of 10.92 and 10.26 hours, respectively. ID~74 is a BD of spectral type M8 and shows variability with the period of 11.09 hours. ID~115 (an M4.5 dwarf) and ID~123 (an M5 dwarf) show a period of 7.98 hours and 9.44 hours, respectively. It was reported that ID~123 has a thick disk from previous studies (Lada et al. \cite{Lada2006}) which can cause magnetically channelled accretion from the disk to the star surface. It can cause large-amplitude aperiodic variation which can mask any periodic variation underneath. ID~181 (a K7 dwarf) shows a period of 9.93 hours.  ID~322 is a relatively faint M7-type BD and shows a periodicity of 4.24 hours with an amplitude of 10 mmag. The faintness of ID~322 may be due to its location behind the high extinction region of the cloud.

In the case of IDs~4, 9, 25, 148, 163, 172, 176, 233, and 323, no periodic nature is detected from our data, but these are aperiodic variables having significant variability in their light curves (see Figure \ref{lc0}). ID~25 and ID~233 both have large amplitudes of variation ($\sim$0.2 - 0.5 mag). However, ID~25 does not show any periodic variation in a short time-scale; it is aperiodic with large variation from mean magnitude (see Figure \ref{lc0}). Flaherty et al. (\cite{Flaherty2013}) reported that ID~25 has a period of 12 days. We do not have large coverage in time-series, and the variation in ID~25 appears to be aperiodic in our analysis (see Figure \ref{lc0}. The source ID~233 appears as a flaring star and we have discussed separately in section \ref{flare}. It was reported that ID~233 and ID~25 have a thick disk, whereas ID~163 has anaemic disk structure (Lada et al. \cite{Lada2006}).

By visual inspection, Cody et al. (\cite{Cody2014}) found a few variable sources (object IDs: 25, 29, 118, 123, 163, 172, 176) in IC~348. We found them variables from our available data, expect ID~118.  ID~172 shows a period of 3.86 hours during the first three nights monitoring during 18-20 December 2016 and the phase light curve of ID~172 is shown in Figure \ref{phase}. However, the phase light curve shows no periodic nature by folding the whole range of data with that period. A few of the sources have shown periodic variation in one observing run, while no periodicity is observed in the next run. The reason for such non-persistent flux variation might be due to the evolving weather pattern, change in star-spot coverage (Scholz et al. \cite{Scholz2009}, Cohen et al. \cite{Cohen2004}) or differential rotation at different latitudes of a storm system with respect to cloud features (Radigan et al. \cite{Radigan2012}, Artigau et al. \cite{Artigau2009}).

A few hypotheses are proposed to explain the mechanism of variability in VLMs and BDs. Such as the aperiodic variability in VLMs like ID4, ID9, ID30, ID90, ID124, ID148, ID163, ID172, ID176, ID233, ID323 are possibly due to the non-uniformly distributed accretion "hot spot". Such spots are originated by magnetically channelled accretion from circumstellar disk to the star/BD surface (Joy \cite{Joy1942}; Reipurth et al. \cite{Reipurth2007}, Mayne \& Harries \cite{Mayne2010} and references therein, Alcala et al. \cite{Alcala2014}) and may cause erratic changes in the light curves.

In contrast, periodic variability is caused by rotational modulation of the flux by cool spots on the stellar surface which are also asymmetrically distributed. Such cool spots (Bailer-Jones \& Mundt \cite{Bailer-Jones2001}, Marti{\'n} et al. \cite{Martin2001}) can form in two processes, one where the atmosphere traps the dust condensates with a permanent hole or low-density cloud formation (Ackerman \& Marley \cite{Ackerman2001}). In rapidly rotating stars, condensate clouds form discrete cloud features like holes in the cloud layer. These holes act as ``hotspots" which cause large-amplitude variability (Radigan et al. \cite{Radigan2012}). The other process is a temporary formation of condensates due to low temperature and raining out after some time. However, it would cause quasi-periodic or aperiodic variation in brown dwarfs. Periodic variability may also happen due to magnetic spots (like our sun), but the atmospheres of brown dwarfs are such that they cannot form stable star-spots (Gelino \cite{Gelino2002}). In our sample, the temperatures of all the sources are higher than 2500 K, which is the upper limit for dust condensation (Helling, Woitke \& Thi \cite{Helling2008}). So there is a high probability that the detected flux variation might be due to cool or hot magnetic spots in a uniform atmosphere with temperature heterogeneities (Radigan et al. \cite{Radigan2012}) and this is corroborated by the fact that magnetic activity is a significant feature in M spectral type (Scholz et al. \cite{Scholz2009}).

However, aperiodic variability can mask low amplitude periodic variation present in the data. The aperiodicity can also be interpreted as the evolution of surface features (dust spots/clouds). Dynamical processes in the cloud structures can cause differential rotation in the atmosphere which can appear as aperiodic changes in short term monitoring (Apai et al. \cite{Apai2013}). This kind of activity is observed in Neptune; there are long term periodic motions which result from clouds and many more small variability features on top of the main periodic signal (Simon et al. \cite{Simon2016}).  To assess the validity of such theories for variability in BDs, further investigation in radial velocity measurements would be useful.

%
%
%

\renewcommand{\tabcolsep}{5pt} 
\begin{sidewaystable*}
\begin{minipage}{140mm}

\vspace*{15.5cm}      
\caption{Details of the identified variables in IC~348}                                                 
\label{table2}
\begin{tabular}{p{0.8cm} p{.8cm} p{4.5cm} p{2cm} p{2cm} p{1.2cm} p{1.2cm} p{1.2cm} p{1.2cm} p{1cm} p{1.7cm} p{1.2cm}}

\hline\\
Star ID & LRL seq. & Identifier & RA (hr:min:sec) & DEC (deg:min:sec) & $Temp^{b,c,h}$ (K)  & J $mag^{a}$    & I $mag^{b,e,i}$  & $SpT^{b}$ & Period (days)  & Period (hours) (this work) & RMS (mag) \\ \hline\\

1 & 1679 & 2MASS J03445205+3158252  & 03 44 52.05   & +31 58 25.2    & 3342   & 13.796  & 16.86  & $M3.5^{f}$  & -  &	3.52 & 0.091 \\ 

3 &1881 & 2MASS J03443379+3158302  & 03:44:33.64 & +31:58:29.1 & 3270 & 13.488 & 16.04 & $M3.75^{f}$& $3.9^{c}$ & 5.72 &  0.026 \\ 

4 &  &   [AMB2013] CFHT-IC 348 17  & 03:44:49.43 & +31:58:44.3 & -  & $15.598^{g}$  & 17.91  &  -  &  - & - & 0.041 \\ 

9 & 363 & 2MASS J03441726+3200152 & 03:44:17.27 & +32:00:15.4 &    2600  & 14.92 & 17.95&   M8 &  -  & - &0.058  \\ 

12& 205 & 2MASS J03442980+3200545     &  	03:44:29.80 & +32:00:54.6  & 2840 &   13.668      & 16.66 & $M6^{j}$	& & 10.92 & 0.014\\ 

25 & 140 & 2MASS J03443568+3203035 & 03 44 35.68 & +32 03 03.5 &    3380  & 14.128 & 17.36 &   M3.25 & $12^{c}$  & -  &0.070 \\ 

29 & 300  & 2MASS J03443896+3203196 & 03 44 38.98 & +32 03 19.8 &    2990  & 14.127 & 17.24 &   M5 &   -    & 10.26  & 0.016 \\ 

30 & 336 & 2MASS J03443237+3203274  & 03 44 32.37 & +32 03 27.4 &2925 &14.884 &17.63 & M5.5  & $1.6^{k}$& - & 0.027\\ 

74 & 405 & [BNM2013] 32.03 1131  & 03:44:21.06 & +32:06:16.1 & 2600 & $15.136^{l}$  & 18.34  & $M8^{b}$   & - & 11.09  & 0.068  \\ 

90 & 291 & Cl* IC 348 LRL 291  & 03 44 34.05 & +32 06 56.9 &    2838  & $13.977^{k}$ & 17.09 &   M7.25 &   $3.4^{c}$  &  -  & 0.041  \\ 

115 & 252 & 2MASS J03442912+3207573 & 03:44:29.11 &	+32:07:51.0 & 3198 	&14.129 & 17.33 & M4.5 &- & 8.00 & 0.022\\

123 & 103 & 2MASS J03444458+3208125 & 03 44 44.58 & +32 08 12.5 &    3560  & 12.881 & 15.49 &   M5 &  -  & 9.44      &0.029\\ 

124 & 355 & 2MASS J03443920+3208136 & 03:44:39.20 & +32:08:13.6 &    2600  & 14.946 & 18.17 &   M8 &    -      & - & 0.047 \\ 

148 &329 & 2MASS J03441558+3209218 & 03:44:15.58 & +32:09:21.8 &  2795   & 14.59  & 17.55 &   M7.5 &   $0.6^{c}$   & - & 0.024\\ 

163 & 182  & 2MASS J03441820+3209593 & 03 44 18.20 & +32 09 59.3 &  3488    & 13.215  & 15.73 &   M4.25  & $2.7^{d}$ &- & 0.130\\ 

172 & 116 & 2MASS J03442155+3210174  & 03 44 21.56 & +32 10 17.4 & 3632 & 12.620  & 15.28  &   M1.5 & $7.0^{c}$   & -& 0.058\\ 

176 & 75 & 2MASS J03444376+3210304 & 03 44 43.76 & +32 10 30.4 & - & 12.294  & 14.91  &   M1.25 & $10.6^{c}$  &- & 0.089\\ 

181 & 41 &  2MASS J03442161+3210376 & 03:44:21.61 & +32:10:37.7 & 4060  & 12.490  & 14.99  & $K7^{b}$    & $2.8^{c}$ & 9.93 &0.025 \\ 

233 & 157 & 2MASS J03441857+3212530 &  03:44:18.58 & +32:12:53.1 &    3451  & 13.816 & 15.92 &   M2.75 &   -      &   - & 0.087 \\ 

322 & 437 & 2MASS J03435638+3209591  & 03:43:56.39 & +32:09:59.1 & 2838   & 15.473  & 18.61 & $M7.25^{b}$   & -  & 4.24 & 0.0105 \\ 

323  & 478 & 2MASS J03443593+3211175   & 03:44:35.94 & +32:11:17.5 & 2810   & 15.890  & 18.52 & $M6.25^{h}$  & - & - & 0.088 \\ 

327 & & Cl* IC 348 LNB 298   & 03:44:39.92 & +32:06:12.87 & - & 15.9   &     19.02     &  -   & -   &     8.63    & 0.100 \\ \hline

\end{tabular}
\end{minipage}
\begin{tablenotes}
\item$^a$2MASS: Cutri et al. \cite{Cutri2003};  $^b$Luhman et al. \cite{Luhman2003};  $^c$Flaherty et al. \cite{Flaherty2013};   $^d$Cohen et al. \cite{Cohen2004}; $^e$Monet et al. \cite{Monet2003} (The USNO-B1.0 Catalog);  $^f$Muench et al. \cite{Muench2007};  $^g$NOMAD Catalog (Zacharias et al. \cite{Zacharias2004});  $^h$Luhman (\cite{Luhman1999});  $^i$Littlefair et al. \cite{Littlefair2005};  $^j$Currie T. \& Kenyon S.J. \cite{CK09};  $^k$Alexander et al. \cite{Alexander2012};  $^l$UKIDSS-DR8 (Lawrence et al. \cite{Lawrence2007}).\\
For simplicity, star ids are utilized in the text more often.\\
The estimated periods have systemic errors. We discussed it in the subsection \ref{periodogram}.
\end{tablenotes}

\end{sidewaystable*}

\subsection{Flare in an young M~2.75 dwarf}
\label{flare}

\begin{figure}[]
\hskip-1cm
\includegraphics[height=2in,width=4in]{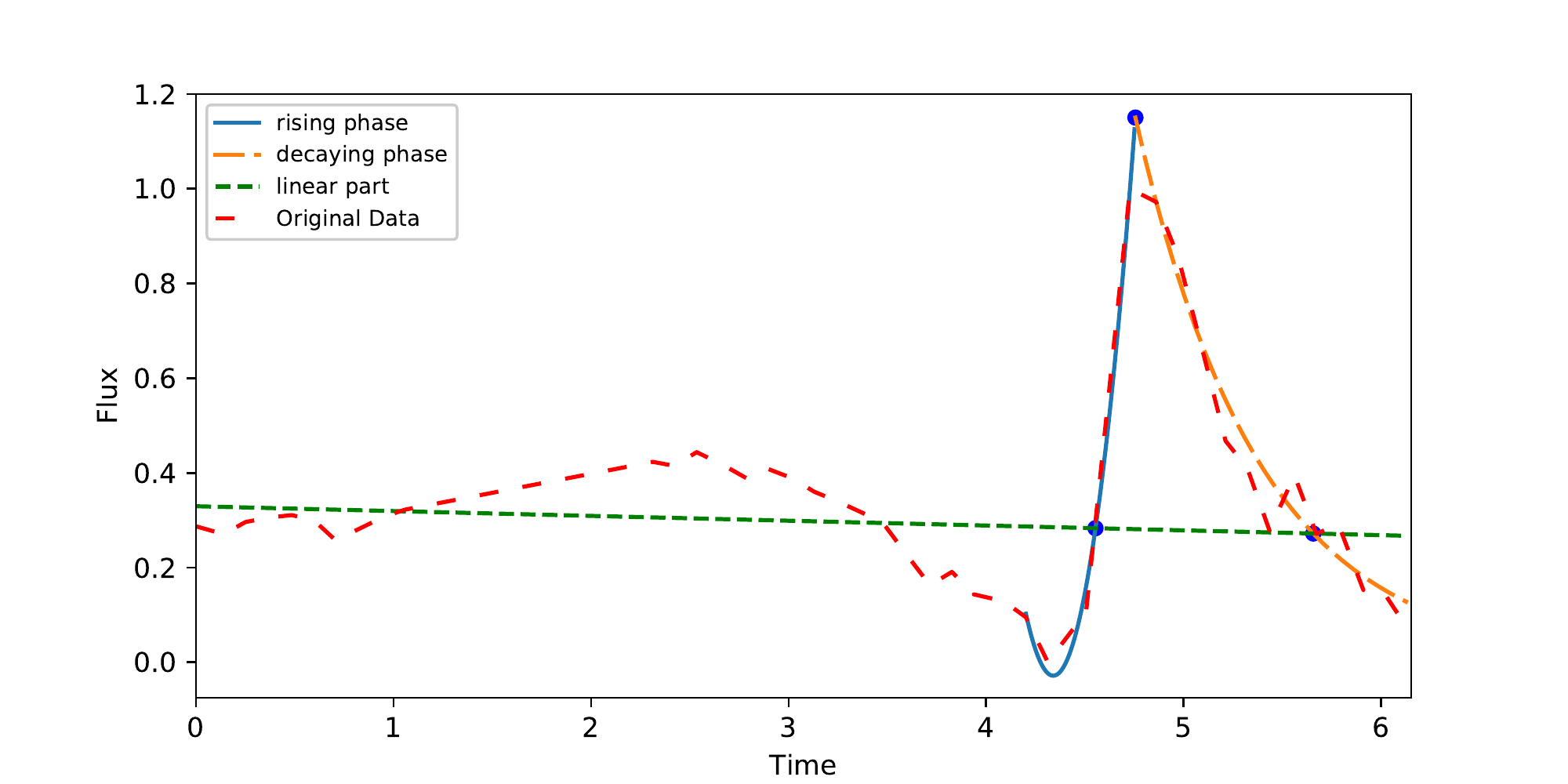}
\caption{A flare event in M2.75 dwarf (ID233) is observed on 20 December 2016. The light curve is fitted in three parts for three different region: Constant Phase (before flare), Rising Phase and Decaying Phase. Three blue triangles are indicating flare starting, peak and ending time in our calculation.}
\label{flare233}
\end{figure}

 We detected an optical flare event from an active M~2.75 dwarf (ID~233) during our observing run on 20 December 2016. In Fig.\ref{flare233}, we have shown the flare light curve of that object, which is a classical flare having only one peak with a sharp rise and fast decay followed by a slower exponential decay (Hawley et al. \cite{Hawley2014}). We have noticed a dimming flux in the pre-flare state just before the rising part in Fig.\ref{flare233}. The pre-flare dimming has been reported previously (e.g., in Hawley et al. \cite{Hawley1995}), and has been explained by a temporary rising of Balmer continuum absorption in chromosphere (Abbett et al. \cite{Abbett1999} \& Allred et al. \cite{Allred2006}).

The flare light curve data provides the flare amplitude (fraction flux, $\Delta F/F$), rise and decay times, duration, and equivalent duration as mentioned in Hawley et al. (\cite{Hawley2014}). Following Davenport et al. (\cite{Davenport2014}) and Gizis et al. (\cite{Gizis2017}), the rising part of the flare light curve is fitted with second-order polynomial, while the fast and slow decaying phase of that is fitted with an exponent function. It is to be noted that due to long cadence in our observations, we might have missed the actual maximum value and the steeper decay part of the light curve, so one component decaying exponent function fits well in the decaying part. The fitting of the flare light curve is shown in Fig.\ref{flare233}. The quiescent flux is estimated from the local mean flux level at the beginning part before the rising and end after the decaying part neglecting the dimming part before the rise, as shown in Fig.\ref{flare233}. If we take $F_{o}$ as the local mean flux and $F_{i}$ as the maximum flux at the flare, then the flare amplitude is defined as $\Delta F/F = (F_i - F_o)/F_o$ (Hawley et al. \cite{Hawley2014}). The estimated flare amplitude, rise and decay times, duration, and equivalent duration from the light curve are listed in Table \ref{table_flare}.

Following Shibayama et al. (\cite{Shibayama2013}) and Yang et al. (\cite{Yang2017}), we have estimated the total energy of the flare event using the stellar luminosity, flare amplitude, and duration of the flare. If we assume that the star is a blackbody radiator of effective temperature (T$_{eff}$), and the observed flare continuum can be described by a blackbody of an effective temperature of 10000 K (T$_{\rm flare}$). Hawley \& Fisher (\cite{Hawley1992}) found that a flared continuum of an M dwarf AD Leo could be described by a blackbody temperature $\sim$8500 - 9500 K in the wavelength range 1000 - 9000 $A^o$. Such estimation may have an error of a few tens of per cent due to clarity of the flare continuum.

The bolometric flare luminosity ($L_{\rm flare}$) could be estimated from $T_{\rm flare}$ and the area of flare ($A_{\rm flare}$) from the following equation,\\
\begin{equation}
~~~~~~~~~~~~~~~~ L_{\rm flare} = \sigma_{\rm SB} T_{\rm flare}^4 A_{\rm flare}~,
\end{equation}

Where ${\bf \sigma_{SB}}$ = Stefan-Boltzmann constant.

For the estimate of  ${\bf A_{\rm flare}}$, we use observed luminosity of star (${\bf L'_{\rm star}}$), flare (${\bf L'_{\rm flare}}$) and flare amplitude of the light curve (${\bf C'_{\rm flare}}$).

\begin{eqnarray}
L'_{\rm star} &=& \int R_{\lambda}B_{\lambda(T_{\rm eff})}d\lambda \cdot \pi R_{\rm star}^2~, \\
L'_{\rm flare} &=& \int R_{\lambda}B_{\lambda(T_{\rm flare})}d\lambda \cdot A_{\rm flare}~, and\\
C'_{\rm flare} &=& \frac{ L^{'}_{flare} }{ L^{'}_{star} } = \frac {F_{i}-F_{0}} {F_{0}}~,
\end{eqnarray}

where $\lambda$ is the wavelength, ${\bf B_{\lambda(T)}}$ is the Plank function, and ${\bf R_{\lambda}}$ is the response function of the 2-m HCT. We can estimate ${\bf A_{\rm flare}}$ from these equation as follows,

\begin{equation}
~~~~~~~~~~~~~~~~ A_{flare} = C_{flare}^{'} \pi R^{2} \frac{\int R_{\lambda} B_{\lambda_{T_{eff}}} d\lambda}{\int R_{\lambda}B_{\lambda_{T_{flare}}} d\lambda}
\end{equation}

${\bf L_{\rm flare}}$ can be estimated from equations (1) and (5), and ${\bf C'_{\rm flare}}$ is a function of time, therefore, ${\bf L_{\rm flare}}$ is also a function of time. Total bolometric energy of the flare (${\bf E_{\rm flare}}$) is an integral of ${\bf L_{\rm flare}}$ during the flare duration,

\begin{equation}
E_{\rm flare} = \int_{\rm flare} L_{\rm flare}(t) dt~.
\end{equation}

We have constructed the Spectral Energy Distribution (SED) for this source with the model based on Robitaille et al. \cite{RT2007} using near-IR (JHK) data from 2MASS point source catalogue (Cutri et al. \cite{Cutri2003}), IRAC 3.6$\mu$m, 4.5$\mu$m, 5.8$\mu$m and 8.0 $\mu$m data (Infrared Array Camera; Fazio et al. \cite{Fazio2004}), and MIPS 24 $\mu$m data (Mid-Infrared  Photometer for Spitzer;  Rieke et al. \cite{Rieke2004}) from Spitzer survey of young stellar clusters (Gutermuth et al. \cite{Gutermuth2009}).   From the SED, the effective temperature, luminosity, and radius of the source are obtained (see Table \ref{table_flare}), and those parameters were used in the estimation of the flared energy.

Using Eqn. (6), we have estimated the flared energy of 1.314$\times10^{35}$ ergs. However, towards this IC~348 source, there is an extinction of $A_v \sim 4.6$ (Lada et al. \cite{Lada2006}). Using IDL's {\it CCM\_UNRED} package, after reddening correction, we get the flare energy,  $E_{\rm flare}  = 1.48\times 10^{35} ergs$. If we take the flare temperatures as 9000 K and 8500 K, then the energies will be $\approx 66 \%$ and $\approx 52 \%$ of the estimated energy, so a few of ten per cent error is associated with the energy calculation.

Flare events are thought to be analogous to the solar flares (Davenport et al. \cite{Davenport2014}). Flares form due to magnetic reconnection events. Due to their turbulent magnetic dynamos, M dwarfs flares are frequent ($\geq 10 \%$ of M dwarfs have flares, Kepler archival database: Yang et al. \cite{Yang2017}, Balona \cite{Balona2015}) and notorious in energy (flare energy to the total stellar energy $\sim 10^{-8} - 10^{-4}$: Yang et al. \cite{Yang2017}). Rapidly rotating young stars with large starspots can produce this kind of superflares, but the frequency of such events is low (Barnes \cite{Barnes2003}, Shibayama et al. \cite{Shibayama2013}). The flare activity depends on the size of the starspots and a small perturbation in chromosphere activity can increase the flare energy to the that of superflare ranges without any other excitation mechanism (Yang et al. \cite{Yang2017}). The rotation period correlates to the chromospheric activity of a star which means a faster rotation period means higher magnetic activity in the object (Pallavicini et al. \cite{Pallavicini1981}). The enormous magnetic energy required to fuel this kind of flare energy can be produced if differential rotation is present at the base of the convection zone (Shibata et al. \cite{Shibata2013}) of the star. These kinds of superflares have a strong effect on the habitability of planets around M-type dwarf stars.

\begin{table}
    \centering
    \caption{Key parameters of the Flaring star }
    \label{table_flare} 
    \begin{tabular}{ p{3cm} p{2.5cm} p{1.4cm}}
    \hline
    Parameter & Value & Reference \\\hline    
  	Object &  2MASS J03441857 +3212530 & 2MASS  \\
    RA [hr:min:secs] & 03:44:18.579 & '' \\
    Dec [deg:min:sec] & +32:12:53.08 & '' \\ 
    $J$ [mag] & 13.816 $\pm$ 0.025 & '' \\
    $I$ [mag] & 15.924 $\pm$ 0.015 & Lt05 \\
    SpT & M2.75 & L03 \\
    T$_{\rm eff}$ [K]    & $2800\pm1050$     & This work \\
    $L/L_{\odot}$ & $0.279\pm2.136$ & '' \\
    R [$R_{\odot}$] & $2.23\pm0.57$ & ''	\\
    E.D. [sec] & 3670.2 & ''\\
    Duration of flare [min] & 68 & '' \\
    Rising Phase [min] & 12  & '' \\
    Decaying Phase [min] & 56  & '' \\ 
    $C'_{\rm flare}$  & 0.1753 &  '' \\
    ${\bf A_{\rm flare}}$ [$m^2$] & 9.4$\times 10^{15}$  & '' \\
	${\bf L_{\rm flare}}$ [erg/s] & 5.34$\times 10^{31}$  & ''	 \\
	$T_{flare}$ [k] & 10000 & '' \\
    $A_v$ & 4.6 & La06  \\
    ${\bf E_{\rm flare}}$ [erg] & 1.48 $\times 10^{35}$  & This work	 \\
    \hline
    
    \end{tabular}
   2MASS : 2MASS (Cutri et al. \cite{Cutri2003});  L03: Luhman et al. \cite{Luhman2003};  Lt05: Littlefair et al. \cite{Littlefair2005}; La06: Lada et al. \cite{Lada2006}
\end{table}

\begin{figure}[]
\hskip-1cm
\includegraphics[height=2.5in]{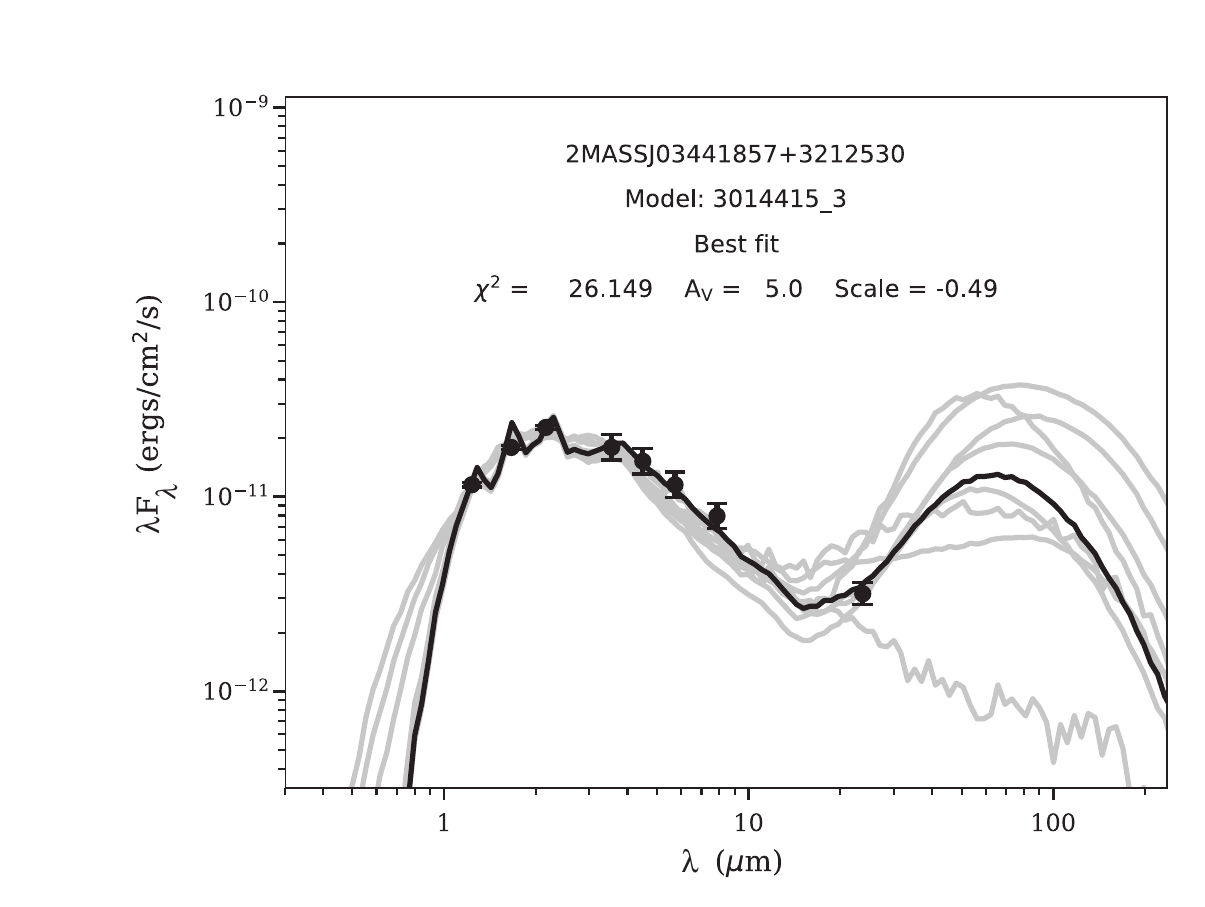}
\caption{Spectral Energy Distribution (SED) of a M2.75 flare star is constructed using the radiative transfer models provided by Robitaille et al. \citep{RT2007}. The grey plots are for equally good fits of the source.}
\label{img_sed}
\end{figure}

\section{Summary and Conclusions}
\label{summary}

In this paper, we presented {\it $I$}-band time-series photometry of well-characterized young VLMs and BDs in the spectral type of M dwarfs, which are bonafide members of IC~348. We summarized the main results as follows:

\begin{enumerate}

\item The {\it $I$}-band (down to $\sim$19 mag) light-curve analysis enables
us to probe low-amplitude variability in young VLMs and BDs of IC~348. From a sample of 177 light-curves of M dwarfs using our new {\it $I$}-band observations, we detect new photometric variability in 22 objects including 6 BDs.

\item Using Lomb-Scargle periodogram analysis, we found that among 22, 11 M dwarfs including a BD show an hour-scale periodic variability in the period range 3.5 - 11 hours, while rest are aperiodic in nature. 

\item Interestingly, an optical flare is detected in a young M2.75 dwarf in one night data on 20 December 2016. From the flare light curve and our constructed SED, we estimate the emitted flared energy of $1.48\times 10^{35}$ ergs. The observed flared energy with an uncertainty of tens of per cent is close to the super-flare range ($\sim$  10$^{34}$ ergs), which is rarely observed in active M dwarfs. 

\item Periodic variability in such low-mass objects is caused by rotational modulation of the stellar flux by an asymmetric distribution of cool spots or spot groups on the stellar surface. While aperiodic variations are probably caused by variable, magnetically affected accretion from the circumstellar disk onto the star.

\end{enumerate}

\section*{Acknowledgements}
This research work is supported by the S N Bose National Centre for Basic Sciences under the Department of Science and Technology, Govt. of India. The authors are thankful to the JTAC members and the staff of the 1.3-m Devasthal optical telescope operated by the Aryabhatta Research Institute of Observational Sciences (ARIES, Nainital), the HTAC members and the staff of the HCT, operated by the Indian Institute of Astrophysics (IIA, Bangalore). SG is grateful to the Department of Science and Technology (DST), Govt. of India for their INSPIRE Fellowship scheme.

\section*{Data Availability}
Data were obtained using the 1.3-m Devasthal Fast Optical Telescope located at Devasthal, Nainital, India (Sagar et al. \cite{Sagar2011}) and the 2-m Himalayan Chandra Telescope, located at Hanle, Ladakh, India. Data are not publicly available but will be provided upon request.







\appendix

\bsp	
\label{lastpage}
\end{document}